\documentstyle[aps,preprint,epsfig]{revtex}

\begin{document}

\preprint{} \draft

\title{Hot electron relaxation: Exact solution for a many electron
  model}

\author{ K.\ Sch\"onhammer and C.\ W\"ohler}

\address{
  Institut f\"ur Theoretische Physik der Universit\"at G\"ottingen,\\
  Bunsenstr.9, D-37073 G\"ottingen, Germany}

\maketitle

\begin{abstract}
  
  The exact nonequilibrium time evolution of the momentum distribution
  for a finite many particle system in one dimension with a linear
  energy dispersion coupled to optical phonons is presented. For
  distinguishable particles the influence function of the phonon bath
  can be evaluated also for a finite particle density in the
  thermodynamic limit. In the case of fermions the exact fulfillment
  of the Pauli principle involves a sum over permutations of the
  electrons and the numerical evaluation is restricted to a finite
  number of electrons. In the dynamics the antisymmetry of the
  wavefunction shows up in the obvious Pauli blocking of momentum
  states as well as more subtle interference effects.  The model shows
  the expected physical features known from approximate treatments of
  more realistic models for the relaxation in the energy regime far
  from the bottom of the conduction band and provides an excellent
  testing ground for quantum kinetic equations.

\end{abstract}


\narrowtext

\section{Introduction}

\label{intro}

The short time evolution of the relaxation of hot electrons due to the
emission of optical phonons is not properly described using a
Boltzmann equation, which assumes energy conserving independent
scattering events, but requires the use of quantum kinetic equations
\cite{zimmerpss159,haugprb47,kuhnprb50,rokuhnprb53}. Uncontrolled
approximations involved in the derivation of these equations can lead
to unphysical results such as negative probabilities for the momentum
distribution. For the testing of new approximations it is very useful
to find exactly solvable models, which show the main expected
properties. We recently proposed two one-dimensional models for the
relaxation of a {\it single} electron in which the energy dispersion
of the excited electron is linearized around the initial energy high
in the conduction band. In the first model \cite{mwfs} an infinite
Fermi sea, as in the Tomonaga-Luttinger-model \cite{mattisjmp6} is
included. This can be viewed as a crude model for the relaxation in a
doped semiconductor , while the second model without the Fermi sea
\cite{mfws} is more relevant for undoped semiconductors. {\it Exact}
results for the momentum distribution were presented and compared with
various improved quantum kinetic equations.

In a typical experimental situation a finite density of electrons is
optically excited into the valence band. The nonequilibrium momentum
distribution can be inferred from energy resolved measurements of
transmission changes $\Delta T / T$ with delay times in the
femtosecond range \cite{furst}. For their theoretical discussion of
the results for GaAs these authors have compared approximate
calculations in more realistic band structures with our second model
\cite{mfws} and ``find the 1D linear model to be well justified as long
as the carriers do not reach the bottom of the band''. For a more
detailed study of the relaxation process it is necessary to treat the
two effects properly which are not correctly described if the results
of the single electron model are simply superposed according to the
density of the excited electrons. 
A phonon emitted by one
electron can be reabsorbed by another one leading to energy gain
satellites in the momentum distribution. In the relaxation of many
electrons the Pauli principle has to be taken into account. In a
Boltzmann equation or typical quantum kinetic equations the Pauli
blocking of states is only treated in an {\it average} way and no
exact description has been provided previously. In this paper we
generalize our second model \cite{mfws} to the {\it many} electron
case and again obtain the exact solution for the nonequilibrium
momentum distribution.

In Sec. \ref{modelsol} we present the model and its solution for two
cases. We first assume the ``electrons'' to be distinguishable. Under
this assumption the results for finite systems with a finite number of
particles can be extended to the thermodynamic limit with a finite
{\it density} of particles. The results show the importance of the
gain processes mentioned earlier. If the fermionic character of the
``electrons'' is properly taken into account, the result for the
momentum distribution contains a sum over permutations of $N-1$
electrons and a numerical evaluation of the exact result is only
possible for finite systems. For not too large electron-phonon
coupling we also present an excellent approximation to the exact
results which can be evaluated for larger systems. The model allows a
detailed study in which parameter regimes a proper treatment of the
Pauli blocking is of importance. Interference effects which are not
correctly described by the usual quantum kinetic equations are shown
to be especially important in the gain region.  Some analytical and
numerical details of the solution as well as a comparison with time
dependent perturbation theory are presented in appendices.

\section{The model and its solution}
\label{modelsol}
In generalization of the polaron model presented in \cite{mfws} we
study a one-dimensional model of the relaxation of $N$ ``hot''
particles due to the interaction with optical phonons. In order to
obtain the exact solution of the time dependent Schr\"odinger equation
we assume a {\it linear} energy dispersion for the particles. The
Hamiltonian for a finite system of length $L$ with periodic boundary
conditions reads

\begin{equation}
  \label{hamiltonian}
  H=H_{e}+H_{p}+H_{ep} = H_{0}+H_{ep}
\end{equation}

\noindent with

\parbox{6.0cm} {\begin{eqnarray*} H_e &=& v\sum_{i=1}^{N} \hat{p_i}
    \;\; , \;\;
    H_p = \sum_{q} \omega_q b_q^{\dagger} b_q^{} \;\; , \\
    H_{ep} & = & \left( \frac{2 \pi}{L} \right)^{1/2} \sum_{ q} g(q)
    \left( \rho_q^{\dagger} b_{q}^{} + b_q^{\dagger} \rho_q^{} \right)
    \;\; ,
  \end{eqnarray*}}
\hfill \parbox{2cm} {\begin{equation} \label{hamiltonianterms}
  \end{equation}}

\noindent where $v$ is the velocity of the particles and
$b_{q}^{\dagger}$ the creation operator of a phonon with frequency
$\omega_{q}$. All the numerical results presented in Secs. \ref{dires}
and \ref{indires} are for optical phonons with a constant phonon
frequency $\omega_q \equiv \omega_0$.  The ``electron-phonon''
coupling strength given by $g(q)$ is specified later and the operator
$\rho_{q}$ denotes the Fourier transform of the particle density

\begin{equation}
  \label{rho}
  \rho_q=\sum_{i=1}^{N} e^{-iq\hat{x_i}}.
\end{equation}

We will discuss the case of distinguishable particles as well as the
model with $N$ spinless fermions. Due to the linear dispersion the
model has no ground state. In the fermionic case this could be altered
by adding an infinite Dirac sea. This is {\it not} considered in the
following and we study the relaxation of an initial state with initial
momenta $k_{1}^{0}, \ldots ,k_{N}^{0}$ and the phonons in their
unperturbed ground state or in thermal equilibrium at a finite
temperature.  The exact description of the time dependence is possible
due to the commutation relation $[\hat{x_{i}},H]=iv \openone$, which
leads to

\begin{equation}
  \label{rhot}
  \rho_{q}(t) = e^{iHt}\rho_q e^{-iHt} = e^{-iqvt}\rho_{q}
\end{equation}

\noindent {\it independent} of the coupling strength $g(q)$. The
Heisenberg equation for $b_{q}(t)$ reads

\begin{equation}
  \label{bqe}
  i \frac{d}{dt} b_q(t) = \omega_q b_q (t) + \left( \frac{2\pi}{L}
\right)^{1/2} g(q) \rho_q(t).
\end{equation}

\noindent The solution of this inhomogeneous equation follows with
(\ref{rhot})

\begin{eqnarray}
  \label{bqt}
  b_q(t)= e^{-i\omega_q t} \left[ b_q +  \left( \frac{2\pi}{L} \right)^{1/2}
  g(q) \left( \frac{e^{-i(vq-\omega_q)t} -1}{vq-\omega_q} \right)
  \rho_q \right] \;\; .
\end{eqnarray}
 
\noindent Therefore the change of the expectation value of the 
number of phonons $N_q=b_{q}^{\dagger}b_{q}$ due to the presence of
the particles is determined by the expectation value of
$\rho_{q}^{\dagger} \rho_{q}$ in the initial state

\begin{equation}
  \label{phores}
  \delta \left<N_q\right>(t) = \frac{2\pi}{L} \left| f_q(t) \right|^2
    \left<\rho_{q}^{\dagger} \rho_{q} \right>
\end{equation}

\noindent with

\begin{equation}
  \label{fqts} 
  \left| f_q(t) \right|^2 = g^2(q)
    \left( \frac{\sin{\left[ \left( \omega_q-vq\right) t/2 \right]}}
      {\left( \omega_q -vq \right) /2} \right)^2. 
\end{equation}

\noindent This will be discussed in Secs. \ref{dires} and
\ref{indires}. We first present the method to calculate expectation
values of an operator $A$ acting in the Hilbert space of the
particles, i.\ e.\ $ A = \left| k \right>_{(i)}{} {}_{(i)}\left< k
\right| $, where $ A=\left| k \right>_{(i)}$ is the momentum state in
the Hilbert space of the particle $i$.  For the case of
indistinguishable particles we will consider one-particle operators $A
= \sum_{i=1}^{N} A_{(i)}$. We present the result for expectation
values $\left< A \right>(t)$ for a statistical operator of the initial
state which factorizes into a particle and a phonon part, i.\ e.\ 
$\rho_0= \rho_p \cdot \rho_{ph}$. We work in the interaction
representation and denote the time evolution with the unperturbed
Hamiltonian by $A_D(t)$. Due to the simple form of
$\rho_{q,D}(t)=e^{-iqvt} \rho_{q}$ it is favorable to perform the
trace in the particle Hilbert space using the position states $\left|
  {\bf x} \right> \equiv \left| x_1 \right>_{1} \otimes \cdots \otimes
\left|x_N \right>_{N}$. This yields

\begin{equation}
  \label{at}
  \left< A \right>(t) = \int d{\bf x}^\prime d{\bf x} \left< {\bf
      x}^\prime \left| \rho_{p} \right| {\bf x} \right> Tr_{ph} \left[
    \rho_{ph} \left< {\bf x} \left| \tilde{U}^{\dagger} (t) A_D(t)
    \tilde{U}(t) \right| {\bf x}^\prime \right> \right],
\end{equation}

\noindent where $\tilde{U}(t) \equiv e^{iH_0 t} e^{-iHt}$ is the time
evolution operator in the interaction representation. The proper
projection on the totally antisymmetric states in the case of fermions
is taken care of by the statistical operator $\rho_p$.  In the
following we assume $\rho_p$ to describe a {\it translational
  invariant} initial state, i.\ e.\ $\rho_p$ is diagonal in the
momentum representation. The simple form of $\rho_{q,D}(t)$ implies
that the only electronic operators in $H_{ep,D}(t)$ are the commuting
position operators $\hat{x}_i$.  Therefore $\tilde{U}(t) \left| {\bf
    x}^\prime \right> = \tilde{U}_{{\bf x}^\prime}(t) \left| {\bf
    x}^\prime \right>$, where $\tilde{U}_{{\bf x}'}(t)$ is an operator
in the {\it phonon} Hilbert space only. This allows us to write
Eq.(\ref{at}) in the form

\begin{equation}
  \label{atres}
  \left< A \right>(t) = \int d{\bf x}^\prime d{\bf x} \left< {\bf
      x}^\prime \left| \rho_{p} \right| {\bf x} \right>
    \left< {\bf x} \left| A_D(t)
    \right| {\bf x}^\prime \right> F({\bf x},{\bf x\prime},t),
\end{equation}

\noindent where the ``influence function'' $F({\bf x},{\bf x}^\prime,t)$ is
given by

\begin{equation}
  \label{inflf}
  F({\bf x},{\bf x}^\prime,t) = Tr_{ph} \left[\rho_{ph}
  \tilde{U}_{{\bf x}}^{\dagger} (t) \tilde{U}_{{\bf x}^\prime} (t) \right].
\end{equation}

The differential equation for $\tilde{U}_{{\bf x}} (t)$ reads

\begin{equation}
  \label{udeq}
  i \frac{d}{dt} \tilde{U}_{{\bf x}}(t) = \left( C^{}_{{\bf x}}(t) +
  C^{\dagger}_{{\bf x}}(t) \right) \tilde{U}_{{\bf x}}(t)
\end{equation}

\noindent with

\begin{equation}
  \label{cxt}
  C_{{\bf x}}(t) = \left( \frac{2\pi}{L} \right)^{1/2} \sum_{q} g(q)
  e^{-i(\omega_q-vq)t} \rho_q^{\dagger}({\bf x}) b_q \;\; ,
\end{equation}

\noindent where $\rho_q({{\bf x}})$ is the eigenvalue of the
operator $\rho_q$ in the state $\left| {\bf x} \right>$. Using the
Baker-Hausdorf formula the solution of equation (\ref{udeq}) is
straightforward \cite{mfws}

\begin{equation}
  \label{uxtsol}
  \tilde{U}_{{\bf x}}(t) = \exp{\left\{ -i \int_0^{t} C_{{\bf
  x}}^{\dagger} (t') dt'  \right\}}
  \exp{\left\{ -i \int_0^{t} C_{{\bf x}}(t') dt'  \right\}} a_{{\bf
  x}}(t)  \;\;,
\end{equation}

\noindent where $a_{{\bf x}}(t)$ is given by the $c$-number

\begin{equation}
  \label{axt}
  a_{{\bf x}}(t)
 = \exp{ \left\{ -\frac{2\pi}{L} \sum_{q} \gamma_q(t)
  \left| \rho_q({\bf x}) \right|^2 \right\} }
\end{equation}

\noindent with
 
\begin{equation}
  \label{gam}
  \gamma_q(t) = \frac{g^2(q)}{\left(\omega_q- v q \right)^2}
  \left( 1-i\left(\omega_q-v q\right) t -e^{-i(\omega_q-v q) t}
  \right) \;\; .
\end{equation}

\noindent The trace over the Hilbert space in Eq.(\ref{inflf}) can be
performed if the phonons are initially assumed to be in thermal
equilibrium at temperature $T$. Then we can use $\left< e^A e^B
\right> = e^{\frac{1}{2} \left< A^2 + 2AB + B^2 \right>}$, valid for
canonical expectation values for free bosons and operators $A$ and $B$
{\it linear} in the boson creation and annihilation operators
\cite{mer}. Using Eqs.(\ref{cxt}) and (\ref{uxtsol}) one obtains

\begin{equation}
  \label{inflfres}
  F({\bf x},{\bf x}',t) = \exp { \left\{ \frac{2\pi}{L}  
    \sum_{q} \left| f_q (t) \right|^2 \left[ \rho_{q}^
    {\ast}({\bf x}) \rho_{q}({\bf x}') - 
    n_B(\omega_q) \left| \rho_{q}({\bf x}) -
    \rho_q({\bf x}^{\prime}) \right|^2 \right] \right\} } a_{{\bf
      x}}^{\ast}(t) a_{{\bf x}'}(t),
\end{equation}

\noindent where $n_B(\omega_q) = 1/(exp(\omega_q/k_B T_{ph})-1)$ is
the Bose function.  With this analytical expression for the influence
function the evaluation of expectation values of particle operators is
reduced to integrations in Eq.(\ref{atres}).

We now specialize the operator $\hat{A}$ to the momentum distribution
and begin with distinguishable particles.  For the operator $n_k^{(1)}
\equiv \left| k \right>_{(1)} {}_{(1)}\left< k \right| $ the integral
in (\ref{atres}) simplifies as

\begin{equation}
  \label{nk1ortsd}
  \left< {\bf x} \right| n_k^{(1)} \left| {\bf x}' \right> 
    = \frac{1}{L} e^{ik(x_1-x_1')} \delta
    (\underline{{\bf x}}-\underline{{\bf x}}')
\end{equation}

\noindent contains a delta function in the variables $\underline{{\bf x}} =
(x_2, \ldots ,n_N)$ and $\underline{{\bf x}}'$. Therefore the
influence function $F$ is needed only for $\underline{{\bf x}}' =
\underline{{\bf x}}$.  In the variables $\tilde{x}_i \equiv x_i-x_1'$
one has

\begin{equation}
  \label{rqprqortsd}
  \rho_q^{\ast}(x_1,\underline{{\bf x}})\rho_q(x_1',\underline{{\bf
      x}}) = e^{iq
    \tilde{x}_1} + \sum_{j=2}^{N}\left(e^{iq(\tilde{x}_1 -
    \tilde{x}_j)}+e^{iq\tilde{x}_j}\right) + \sum_{i,j=2}^{N} e^{iq(
    \tilde{x}_i - \tilde{x}_j)}
\end{equation}
 
\noindent and corresponding expressions for $\left| \rho_q (x_1,
  \underline{{\bf x}}) \right|^2$ and $\left| \rho_q
  (x_1',\underline{{\bf x}}) \right|^2$. Due to the identity
$\gamma_q^{\ast}(t) + \gamma_q(t) = \left| f_q(t) \right|^2$ the terms
with the double sum involving two ``other'' particles cancel. This
leads to

\begin{equation}
  \label{inflf1orts}
  F(x_1,\underline{{\bf x}};x_1',\underline{{\bf x}},t) = F_1(\tilde{x}_1,t)
  \prod_{i=2}^{N} \exp{ \left\{ \frac{2\pi}{L} \sum_{q} \left[
    \gamma_q^{\ast}(t) \left(e^{iq \tilde{x}_1}-1 \right) e^{-iq 
      \tilde{x}_i} - \gamma_q(t) \left(e^{-iq \tilde{x}_1}-1 \right) e^{iq
      \tilde{x}_i} \right] \right\} }
\end{equation}

\noindent where 

\begin{equation}
  \label{f1xt}
  F_1(x,t) = \exp{\left\{ \frac{2\pi}{L} \sum_{q} \left|f_q(t)
  \right|^2 \left[ \left(e^{iqx}-1 \right) \left(1+n_B(\omega_q)
\right) + \left(e^{-iqx}-1 \right) n_B(\omega_q) \right] \right\} }
\end{equation}

\noindent is the influence function of a {\it single} particle
\cite{mfws}.

Now the integration in Eq.(\ref{atres}) can be partially carried out
using Eqs.(\ref{nk1ortsd}) and (\ref{inflf1orts}) and one obtains

\begin{equation}
  \label{nk1tres}
  \left< n_k^{(1)} \right>(t) = \int dx e^{ikx} \rho^{(1)}(x) 
    F_1(x,t) \left( G_0(x,t) \right)^{N-1}
\end{equation}

\noindent with

\begin{equation}
  \label{gnxt}
  G_0(x,t)= \frac{1}{L} \int dy \exp{ \left\{ \frac{2\pi}{L} \sum_{q}
    \left[ \gamma_q^{\ast}(t) \left(e^{iqx}-1 \right) e^{-iqy} -
      \gamma_q (t) \left(e^{-iqx}-1 \right) e^{iqy} \right] \right\} }
\end{equation}

and $\rho^{(1)}(x)$ is given by the real space matrix elements of the
reduced statistical operator for particle one for which the trace $(
tr' )$ over the other particles has been performed

\begin{equation}
  \label{statop1x}
  \rho^{(1)}(x_1-x_1') = L^{N-1} \left< x_1,
  \underline{\bf x} \right| \rho_p \left| x_1', \underline{\bf x} 
  \right> = \left< x_1 \right| tr'( \rho_p ) \left| x_1' \right> .
\end{equation}

\noindent For $N=1$ and $\rho^{(1)} = \left| k_1^0
\right> \left< k_1^0 \right|$ , i.\ e.\ $\rho^{(1)}(x) = \frac{1}{L}
e^{-ik_1^0 x}$ Eq.(\ref{nk1tres}) reduces to the polaron result
presented in \cite{mfws}.  The result for $\left< n_k^{(1)} \right>
(t)$ will be analyzed in the next section.

We now turn to the case of $N$ fermions. In first quantization the
momentum occupation number operator $\hat{n}_k$ reads in the Hilbert
space $H_N^{(a)}$, where $(a)$ denotes the subspace of antisymmetrized
states

\begin{equation}
  \label{nkfop}
  \hat{n}_k = \frac{1}{(N-1)!} \sum_{k_2,\ldots,k_N} \left|
  k,k_2,\ldots,k_N \left>_{a}{} _{a}\right< k_N,\ldots,k_2,k \right|
\end{equation}

\noindent The factor $1/(N-1)!$ is necessary because the $N-1$
summations are unrestricted. To calculate $\left< \hat{n}_k \right>
(t)$ we can proceed as in Eqs.(\ref{at}) and (\ref{atres}) using
product states $\left| {\bf x} \right>$ which are {\it not}
antisymmetrized. In the following we describe the time evolution of an
initial state $\left| {\bf k}_0 \right>_a$. The result for an
arbitrary translational invariant statistical operator is obtained by
averaging the result with the normalized probability density $p({\bf
  k}_0)$ describing $\rho_p$.  The initial state $\left|
  k_1^0,\ldots,k_N^0 \right>_a$ and the projectors in Eq.(\ref{nkfop})
take care of the Pauli principle. Using $\left< {\bf x} | {\bf k}
\right>_{a} ={} _{a}\left< {\bf x} | {\bf k} \right>$ we obtain

\begin{equation}
  \label{nkforts}
  \left< {\bf x} \left| \hat{n}_k \right| {\bf x}' \right> =
    \frac{1}{L} \frac{1}{N!(N-1)!} \left[ e^{ik(x_1-x_1')} \delta(
    \underline{{\bf x}} - \underline{{\bf x}}') \pm \cdots \right],
\end{equation}

\noindent where the parenthesis contains all $(N!)^2$ terms which
result from antisymmetrizing the first term with respect to the ${\bf
  x}$ {\it and} the ${\bf x}'$ variables. As in Eq.(\ref{atres}) this
expression is multiplied by the function $f({\bf x},{\bf x}',t) =
\left<{\bf k}^0 | {\bf x} \right>_a {} _{a}\left< {\bf x}' | {\bf k}^0
\right> F({\bf x},{\bf x}',t)$, which is also antisymmetric in both
the ${\bf x}$ and ${\bf x}'$ variables, the ${\bf x}-$ and ${\bf x}'-$
integrations yield $(N!)^2$ identical terms, i.e.\ one can just take
the first term on the rhs of Eq.(\ref{nkforts}) and multiply it by
$(N!)^2$

\begin{equation}
  \label{nkft}
  \left< \hat{n}_k \right> = \frac{N}{L} \int dx_1 dx_1'
    d\underline{{\bf x}} e^{ik(x_1-x_1')}{} _{a}\left< {\bf k}^0 | x_1,
    \underline{{\bf x}} \right> \left< x_1', \underline{{\bf x}} | {\bf k}^0
  \right>_a F(x_1,\underline{{\bf x}};x_1',\underline{{\bf x}},t).
\end{equation}

\noindent Note that again only the special form of the arguments of the
influence function as in Eq.(\ref{inflfres}) appear. Due to the
translational invariance discussed there, the $x_1'-$integration can
be carried out by setting $x_1'=0$ in the remaining integrand and
dropping the factor $1/L$. The factors in the integrand which take
care of the Pauli principle can be written as

\begin{equation}
  \label{instaorts}
  _{a}\left< {\bf k}_0 | x_1,\underline{{\bf x}} \right> \left<
  0, \underline{{\bf x}} | {\bf k}_0 \right>_a = \frac{1}{NL}
  \sum_{i,j=1}^{N} e^{ik_i^0x_1}{} _{a}\left< {\bf k}^0 |
  c_{k_i^0}^{\dagger} | \underline{{\bf x}} \right> \left< 
  \underline{{\bf x}} | c_{k_j^0} | {\bf k}^0 \right>_a
\end{equation}

\noindent where the $c_k^{\dagger} (c_k)$ are the usual creation
(annihilation) operators and the factor $1/N$ appears because we have
switched from antisymmetrized $N-$particles states to antisymmetrized
$(N-1)-$particle states. In Eq.(\ref{nkft}) the rhs of
Eq.(\ref{instaorts}) is integrated with a function {\it symmetric} in
the $\underline{{\bf x}}$ variables. In writing out the factors $_{a}
\left< {\bf k}^0 \right| c_{k_i^0} \left| \underline{{\bf x}} \right>$
and $\left< \underline{{\bf x}} \right| c_{k_j^0} \left| {\bf k}^0
\right>_a$ as antisymmetrical sums of $(N-1)!$ terms each term of the
second sum yields the same contribution. Therefore for the use of the
integration with a symmetric function in $\underline{{\bf x}}$ we can
write

\begin{equation}
  \label{instaortsres}
  _{a}\left< {\bf k}_0 | x_1,\underline{{\bf x}} \right> \left<
  0, \underline{{\bf x}} | {\bf k}_0 \right>_a = \frac{1}{NL^N}
  \sum_{i,j=1}^{N} e^{i(k-k_i^0)x_1} \sum_{P_{N-1}}(-1)^{i+j+P_{N-1}}
  e^{-i\sum_{l=2}^{N} (k_{P_{N-1},l}^{0|i} - k_{l}^{0|j}) x_l}.
\end{equation}

\noindent Here we have introduced the notation $(l=2,\ldots,N)$

\begin{equation}
  \label{klnohnej}
  k_l^{0|j} \equiv \left\{ \begin{array}{ll}
  k_{l-1}^0 & (l \leq j) \\
  k_l^0 & (l>j)
  \end{array} \right. 
\end{equation}
 
\noindent i.e.\ the upper index indicates the missing initial momentum. 
If one inserts (\ref{instaortsres}) on the rhs of Eq.(\ref{nkft}) one
obtains

\begin{equation}
  \label{nktres}
  \left< n_k \right>(t) = \sum_{i=1}^N \frac{1}{L} \int dx e^{i(k-k_i^0)x} 
    F_1(x,t) G_{N-1}^{(i)}(x,t)
\end{equation}

\noindent with

\begin{equation}
  \label{gxt}
  G_{N-1}^{(i)}(x,t) = \sum_{j=1}^{N} \sum_{P_{N-1}} (-1)^{i+j+P_{N-1}}
  \prod_{l=2}^{N} G(k_l^{0|j} - k_{P_{N-1},l}^{0|i},x,t)
\end{equation}

\noindent and

\begin{equation}
  \label{gkxt}
  G(k,x,t) \equiv \frac{1}{L} \int dy e^{iky} \exp { \left\{
    \frac{2\pi}{L} \sum_{q} 
    \left[ \gamma_q^{\ast}(t) \left(e^{iqx}-1 \right) e^{-iqy} -
      \gamma_q (t) \left( e^{-iqx}-1 \right) e^{iqy} \right] \right\} }.
\end{equation}

This is a central result of our paper. The result for $\left< n_k
\right>(t)$ in Eq.(\ref{nktres}) should be compared with the result
for $\left< n_k^{(i)} \right>(t)$ in Eq.(\ref{nk1tres}) summed over
all $N$ particles. If all the initial momenta are different also for
the distinguishable particles the total momentum distribution is given
by Eq.(\ref{nktres}) with $G_{N-1}^{(i)}(x,t)$ replaced by $\left(
  G_0(x,t) \right)^{N-1}$. For fermions the sum over permutations in
Eq.(\ref{gxt}) is necessary in order to obey the Pauli principle
exactly. While for distinguishable particles it is possible to obtain
results for a finite {\it density} of particles in the thermodynamic
limit as discussed in Sec.\ref{dires}, the more complicated expression
for $G_{N-1}^{(i)}(x,t)$ can only be evaluated for a finite number of
particles. The results are presented in Sec. \ref{indires}.

\section{Results for distinguishable particles}

\label{dires}
As we are mainly interested in modelling aspects of hot electron
relaxation in semiconductors, the fermion result Eq.(\ref{nktres}) for
the time dependence of the momentum occupation numbers is considered
more important than the expression Eq.(\ref{nk1tres}) for
distinguishable particles. Due to the relative simplicity of the
latter result compared to the fermionic one, it is useful to discuss
Eq.(\ref{nk1tres}) in order to obtain a first understanding of the
difference between the cases $N=1$ and $N>1$.

The modification of the energy transfer to the phonons is trivial, as
$\left< {\bf k_0} \left| \rho_{q}^{\dagger} \rho_{q} \right| {\bf k_0}
\right> = N$ for $q \neq 0$ , i.\ e.\ the $N=1$ polaron result is just
multiplied by $N$ in Eq.(\ref{phores}). In contrast to this simple
expression, the occupancy $\left<n_{k}^{(1)} \right>(t)$ for $N>1$
differs {\it qualitatively} from the $N=1$ result. With the phonons
initially in their ground state {\it no} energy gain is possible for
one particle, if in our model the $q$-sum in $H$ in
Eq.(\ref{hamiltonian}) is restricted to positive values. If the sum
also includes negative $q$-values a small weight in the gain region
occurs for short times but {\it no} resonant gain peak appears.  In
the following we therefore use two special choices for the coupling
function of the ``electron''-phonon interaction, $g^{(1)}(q)=g
\Theta(q) \Theta(q_c-q)$, where $q_c$ is a cutoff and $g^{2}(q) \equiv
g$ for all $q \in \left( - \infty, \infty \right)$. As discussed in
appendix \ref{exint} the latter choice allows the analytical
calculation of the integrals in the influence function $F({\bf x},{\bf
  x}',t)$ in Eq.(\ref{inflfres}) in the thermodynamic limit. We start
our discussion with the special case $\rho^{(1)} = tr' \rho_p = \left|
  k_1^0 \right> \left< k_1^0 \right|$ , i.e.\ particle one is
initially in the state $\left| k_1^0 \right>$.

A straightforward perturbation expansion of $G_0(x)$ in
Eq.(\ref{gnxt}) in powers of $\gamma \sim g^2$ shows that energy gain
{\it is} possible for $N>1$ in contrast to the $N=1$ result for
$g^{(1)}(q)$ :

\begin{equation}
  \label{gnxtp}
  G_0(x,t) = 1- \left(\frac{2\pi}{L} \right)^2 \sum_q \left\{ \left[ \left|
  \gamma_q(t) \right|^2 + Re \left( \gamma_q(t) \gamma_{-q}(t) \right) \right] 
  \left( 2-e^{iqx}-e^{-iqx} \right) \right\} + O(\gamma^3)
\end{equation}

For the model with $g^{(1)}(q)$ the contribution proportional to
$e^{-iqx}$ leads to a nonzero weight in $\left< n_k^{(1)} \right>(t)$
for $k>k_1^0$, i.e.\ energy gain.  Only the term proportional to $
\left| \gamma_q(t) \right|^2 $ is nonzero and the probability for the
gain process agrees with the result from squaring the amplitude in
second order time dependent perturbation theory for the process in
which particle $2$ emits a phonon, which is reabsorbed by particle $1$
(see appendix \ref{gainpeak}). As $ \left| \gamma_q(t) \right|^2$ is
of the order $(1/L)^0$ the probability of such a gain process goes to
zero as $1/L$ for $L \rightarrow \infty$, if only two (or a finite
number of) particles are present. It is therefore more interesting to
study the thermodynamic limit for a finite (other) particle density
$n_p \equiv (N-1)/L$. In the limit $L \rightarrow \infty$ the momentum
distribution goes over to a function defined on ${\rm I \! R}$.  In
order to normalize this function to one, $\left< n_k^{(1)} \right>
(t)$ has to be multiplied by the density of k-points, i.\ e.\ we
consider the function $ n(k,t) \equiv \lim_{L \rightarrow \infty}
\left( \frac{L}{2\pi} \right) \left<n_k^{(1)} \right>(t) $. As the
function $ G_0(x,t) $ in Eq.(\ref{gnxt}) has the form

\begin{equation}
  \label{gnxtip}
  G_0(x,t) = 1 + \frac{1}{N-1} n_p c_L(x,t),
\end{equation}

\noindent where $c_L(x,t)$ goes over to a well defined limiting
function $c(x,t)$ for $L\rightarrow \infty$, the factor $
\left(G_0(x,t) \right)^{N-1} $ in Eq.(\ref{nk1tres}) has the $ N
\rightarrow \infty$ limit $\exp{ \left( n_p c(x,t) \right) } $. This
yields with the normalized probability $p(k_1)$ for the initial
momentum the exact finite density momentum distribution in the
thermodynamic limit

\begin{equation}
  \label{nktip}
  n(k,t) = \int_{-\infty}^{\infty} dk_1 p(k_1) 
  \int_{-\infty}^{\infty} \frac{dx}{2\pi}
  e^{i(k-k_1)x} e^{b(x,t)+n_p c(x,t)}
\end{equation}

\noindent with

\begin{equation}
  \label{bxttl}
  b(x,t) = \int_{-\infty}^{\infty} dq \left|f_q(t)
      \right|^2 \left[ \left(e^{iqx}-1 \right) \left(1+n_B(\omega_q)
        \right) + \left(e^{-iqx}-1 \right) n_B(\omega_q) \right]
\end{equation}

\noindent and

\begin{equation}
  \label{cixt}
  c(x,t) = \int_{-\infty}^{\infty}dy \left[ \exp{\left\{
    \int_{-\infty}^{\infty}dq \left[ \gamma_q^{\ast}(t) \left(e^{iqx} -1
    \right) e^{-iqy} - \gamma_q(t) \left(e^{-iqx}-1 \right) e^{iqy} \right]
  \right\} } -1 \right]
\end{equation}

\noindent The function $c(x,t)$ cannot be determined completely
analytically even for the coupling function $g^{(2)}(q) \equiv g$, as
discussed in appendix \ref{exint}. It is therefore useful also to
consider $c(x,t)$ in leading order perturbation theory in $\gamma$.
Using Eq.(\ref{gnxtp}) or (\ref{cixt}) one obtains

\begin{equation}
  \label{cxtp}
  c_2 (x,t) = -2\pi \int_{-\infty}^{\infty} dq \left[ \left|
  \gamma_q(t)\right|^2 + Re \left( \gamma_q(t) \gamma_{-q}(t) \right) \right]
  \left( 2-e^{iqx}-e^{-iqx} \right).
\end{equation}

\noindent In this approximation the result for $n^{(1)}(k,t)$ has the
same form as for a {\it single} particle and the phonons at finite
temperature \cite{mfws}, but with the Bose function in
Eq.(\ref{bxttl}) replaced by the sum of $n_B (\omega_q)$ and the time
dependent quantity $ 2\pi n_p \left[ \left| \gamma_q(t) \right|^2 + Re
  \left( \gamma_q(t) \gamma_{-q}(t) \right) \right]/ \left| f_q(t)
\right|^2 $. In order to estimate in which parameter range
Eq.(\ref{cxtp}) provides a good approximation it is useful to
introduce the dimensionless coupling constant $\alpha \equiv 2\pi
g^2/(v\omega_0)$ and to multiply $n_q$ by the relevant length
$v/\omega_0 \equiv 1/q_B$ in our model. If in addition we introduce
dimensionless time and space variables $\tau \equiv \omega_0t$ and $u
\equiv q_Bx$, the finite density contribution $n_pc(x,t)$ in the
exponent in Eq.(\ref{nktip}) can be written as
$(n_p/q_B)\tilde{c}(u,\tau )$ where $\tilde{c}$ is a dimensionless
function. The small density regime is given by $\tilde{n}_p \equiv
n_p/q_B \ll 1$. For finite systems this corresponds to
$(N-1)/(2 \pi n_B) \ll 1$, where $n_B$ is the number of momentum
states in an arbitrary momentum interval $(q,q+q_B)$. The use of
$c^{(2)}(x,t)$ in Eq.(\ref{nktip}) is certainly allowed in the
low-density and small $\alpha$ regime.

For the special case of a sharp initial momentum $p(k_1) = \delta (k_1
- k_1^0)$ the momentum distribution $n(k,t)$ in Eq.(\ref{nktip})
consists of a delta peak at $k=k_1^0$ with a weight $p_{k_1^0} (t) =
\exp \left( b(\infty,t) + n_p c(\infty,t) \right)$ and a continuous
part which for finite times is nonzero for $k_1^0$ also. In
Fig.\ref{tdexst} we compare the time evolution of the weight
$p_{k_1^0} (t)$ for the two different values of the dimensionless
density $\tilde{n}_p = 10^{-3}$ and $\tilde{n}_p = 10^{-2}$ with the
$N=1$ polaron result. The dimensionless coupling constant has the
small value $\alpha = 2 \pi 10^{-3}$. The result for the coupling
functions $g^{(1)}$ and $g^{(2)}$ do not differ on the scale of the
figure. Obviously the presence of the other particles accelerates the
relaxation process. In order to obtain a simple quantitative estimate
of this effect we compare the exact result with the approximation to
replace $c(\infty,t)$ by $c_2(\infty,t)$, where $c_2(\infty ,t)$ is
obtained by dropping the $x-$dependent part on the rhs of
Eq.(\ref{cxtp}). In this approximation the exponential decay of the
polaron case is replaced by the faster decay approximately given by
$p_{k_1^0}(\tau) \approx \exp(-\alpha \tau - \frac{2}{3} \ \alpha^2
\tilde{n}_p \tau^3)$, where we have used Eq.(\ref{pk10tau}). For
dimensionless times $\tau$ larger than $(\alpha \tilde{n}_p)^{-1/2}$
the deviation from the polaron result becomes important.

In Fig.\ref{distl50} the first two curves represent the {\it
  continuous} part of the momentum distribution as a function of the
dimensionless momentum variable $\tilde{k} \equiv k/q_B$ for $k_1^0 =
0$ and the weight of the delta peak is given in the figure captions.
We compare the results for the coupling functions $g^{(1)}$ and
$g^{(2)}$ for the same values of the coupling $\alpha$ as in
Fig.\ref{tdexst} and $\tilde{n}_p = 10^{-3}$ at the time $\tau = 50$
where the influence of the other particles in the energy loss regime
at negative momenta is still rather small. The momentum distribution
shows loss features around $\tilde{k} = -1, -2$ and $-3$ corresponding
to the emission of one, two and three phonons. The short time energy
uncertainty well known from standard derivations of Fermi's golden
rule is clearly visible, especially at the tails of the one phonon
loss peak. For longer times these oscillations weaken and the peaks
become narrower at the resonant positions. A rather small energy gain
peak at $\tilde{k} = 1$ is also visible in Fig.\ref{distl50}. A
qualitative difference of the coupling functions $g^{(1)}$ and
$g^{(2)}$ shows up at positive momenta. While the weight is strictly
zero for $\tilde{n}_p = 0$ in this range for $g^{(1)}(q)$, the
coupling $g^{(2)}(q)$ shows a weak oscillatory weight there as in the
polaron case which survives in Fig.\ref{distl50}. Within the plotting
accuracy the results in this figure agree with the exact results, when
$c(x,t)$ is approximated by $c_2(x,t)$ (Eq.(\ref{cxtp})). For much
larger times many sharp phonon replicas develop and the oscillations
in the tails become weak. The difference between the $g^{(1)}$ and
$g^{(2)}$ results become very small and the replacement of $c(x,t)$ by
$c_2(x,t)$ is quantitatively no longer sufficient especially in the
gain region.
The oscillatory behaviour of both systems is smeared out and the peaks
are broadened if a Gaussian probability distribution $p(k_1)$ is used
in Eq.(\ref{nktip}).

In Fig.\ref{tdlfins} the momentum distribution functions for the
thermodynamic limit with different particle densities and for a
finite system are compared at the time $\omega_0 t = 30$.  Our method
to perform the integration in Eq.(\ref{nk1tres}) for finite systems is
described in appendix \ref{recursion}.  Only the zero momentum state
was initially excited but due to the larger coupling $\alpha = 2 \pi
10^{-2}$ the curves are smoother and the relaxation faster.  Also, for
the first two systems with the larger particle density the energy
uncertainty oscillations of the first resonant state are smeared out.
They have the same dimensionless particle density, i.e.\ for the
thermodynamic limit it is $n_p/q_B = 0.2/(2 \pi)$ and the finite
systems $n_p/q_B = (N-1)/(2 \pi n_B)$ where $q_B = \frac{2
  \pi}{L} n_B$. As can be seen the result for the finite system
with nine particles is already very close to the result for the
thermodynamic limit. At earlier times the two curves agree very
well but then the differences appear and grow with time advancing.
For a smaller coupling this behaviour is slowed down, e.\ g.\ for
$\alpha = 2 \pi 10^{-3}$ the agreement is still excellent at time
$\omega_0 t = 50$.

Comparing the first and the last system one can see that the gain
satellites at positive momentum and also at momentum less than $-4q_B$
are much more pronounced for the larger density. Generally, a larger
density smoothes the curves and accelerates the spreading of the
peaks.

\section{Results for fermions}

\label{indires}
The role of the Pauli principle for more than one fermion is quite
easily seen for the number of phonons created Eq.(\ref{phores}), while
the momentum distribution Eq.(\ref{nktres}) requires a much larger
numerical effort than in the case of distinguishable particles.

In order to obtain $\delta \left<N_q \right>(t)$ we have to evaluate
the expectation value of $\rho_q^{\dagger}\rho_q$ in the fermionic
initial state, which was taken as the Slater determinant $\left| {\bf
    k}^0 \right>_a$ in Sec. \ref{modelsol}. The expectation value can
either be calculated using Wick's theorem using the method of second
quantization or using first quantization as in Sec. \ref{modelsol}.
For $q \neq 0$ one obtains

\begin{eqnarray}
  \label{rdrexp}
  \left< \rho_q^{\dagger} \rho_q \right>_{t=0} & = & \sum_k \left< n_{k+q}
    \right>(t=0) \left(1-\left<n_k\right>(t=0) \right) \nonumber \\
    & = & N - \sum_{i,j=1}^{N} \delta_{k_i^0,k_j^0+q},
\end{eqnarray}

\noindent where the first equality holds for an arbitrary initial
Slater determinant. The second equality shows that the Pauli blocking
on the energy transfer to the phonons is most prominent in the
artificial state in which all the initial momenta are separated by
$q_B$. Then $\left< \rho_{q_B}^{\dagger} \rho_{q_B} \right> = 1$ and
due to the Pauli blocking the number of phonons corresponding to the
energy conserving transitions is reduced drastically compared to the
case of distinguishable particles.

In order to calculate the momentum distribution Eq.(\ref{nktres}) the
functions $G_{N-1}^{(i)}$ have to be evaluated, which involve products
of the functions $G(k,x,t)$ for different values of the momentum
argument. We first discuss the behaviour of $G(k,x,t)$ in leading
order perturbation theory in the electron-phonon coupling $g$.
Expanding the exponential function in Eq.(\ref{gkxt}) we obtain

\begin{eqnarray}
  \label{gkxtp}
  G(k,x,t) & = & \delta_{k,0}+ \left(\frac{2\pi}{L}\right) \left(
    \gamma_k^{\ast} (t) -
    \gamma_{-k}(t) \right) \left( e^{ikx}-1 \right) \nonumber \\
  && - \left( \frac{2 \pi}{L} \right)^2 \sum_{q} \left[ \gamma_{k+q}^{\ast}
    (t) \gamma_q (t) \left( 1+e^{ikx}-e^{i(k+q)x}-e^{-iqx} \right)
  \right. \nonumber \\
  && \left. + \gamma_q^{\ast}(t) \gamma_{k-q}^{\ast}(t) \left(1+ e^{ikx}
      -e^{i(k-q)x} -e^{iqx} \right)/2 \right. \nonumber \\
  && \left. + \gamma_q (t) \gamma_{-k-q}(t)
    \left( 1+ e^{ikx} -e^{i(k+q)x}-e^{-iqx} \right)/2 \right] \nonumber \\
  &&    +O(\gamma^3).
\end{eqnarray}

In the sum over $j$ on the rhs of Eq.(\ref{gxt}) the term
$i=j$ and the terms $j \neq i$ behave differently concerning a
perturbation expansion. We first discuss the diagonal term. In the sum
over the permutations the identity presents the leading approximation
given by $\left( G_0(x,t) \right)^{N-1}$, i.e.\ this contribution to
$G_{N-1}^{(i)} (x,t)$ is identical to the exact result for
distinguishable particles. The contributions of the other permutations
are of the order $(\gamma/L)^m G_0^{N-1-m}$ with $m \geq 2$. For $i
\neq j$ there is no term in which all momentum differences vanish.
The leading term in $\gamma/L$ is given by
$-G(k_i^0-k_j^0,x,t)(G_0(x,t))^{N-2}$. Just keeping the leading order
terms for the diagonal and the non-diagonal terms provides the
following approximation for $G_{N-1}^{(i)}$

\begin{equation}
  \label{gxtp}
  G_{N-1}^{(i)} \approx \left( G_0(x,t) \right)^{N-2} \left( G_0(x,t)-
  \sum_{j (\neq i)} G(k_i^0 - k_j^0,x,t) \right)
\end{equation}

\noindent with an error of order $(\gamma/L)^2$. Therefore 
in this approximation the Pauli principle is strictly obeyed to order
$g^2$. To this order $\left< \hat{n}_k \right>(t)$ in
Eq.(\ref{nktres}) can easily be calculated using $G_0 \approx 1$ and
Eq.(\ref{gkxtp}) in Eq.(\ref{gxtp})

\begin{eqnarray}
  \label{nktp}
  \left< \hat{n}_k \right>^{(1)} (t) &  \approx & \sum_{i} \left\{ 
    \delta_{k,k_i^0} \left( 1-
    \frac{2\pi}{L} \sum_q \left| f_q(t) \right|^2 \right) +
    \frac{2\pi}{L} \left|f_{k_i^0-k}(t) \right|^2 \right\}   \nonumber\\
  &&   + \frac{2\pi}{L} \sum_{i,j} \left| f_{k_i^0-k_j^0}(t) \right|^2
    \left( \delta_{k,k_i^0} - \delta_{k,k_j^0} \right) 
\end{eqnarray}

\noindent where the second line on the rhs via the approximate
$G_{N-1}^{(i)}$ takes care of the Pauli principle. This is most easily
seen for occupancy of an initial momentum state of momentum $k_l^0$

\begin{equation}
  \label{nkltp}
  \left< \hat{n}_{k_l^0} \right>^{(1)}(t) \approx 1 - \frac{2\pi}{L} 
  \sum_{q \neq (k_l^0-k_i^0)} \left|f_q(t) \right|^2.
\end{equation}

The electron with initial momentum $k_l^0$ cannot scatter into the
momentum $k_i^0$ and none of the other electrons can scatter into
momentum $k_l^0$.

In the model with the coupling $g^{(1)}(q)$ gain contributions first
occur in order $g^4$ when the phonons are in their ground state
initially.  For $k$ larger than the initial momenta, i.e.\ $k>k_i^0,
i=1,\ldots N$ it is therefore illuminating to expand the exact
solution to this order in $g$. One can distinguish two contributions
to order $g^4$. First the approximation given in Eq.(\ref{gxtp}) can
be expanded to this order using Eq.(\ref{gnxtp}) and Eq.(\ref{gkxtp}).
The correction to the result for distinguishable particles in
Eq.(\ref{gxtp}) is obtained by using
$(G_0(x,t))^{N-2}G(k_i^0-k_j^0,x,t)\approx G(k_i^0-k_j^0,x,t)$ and
$F_1(x,t)\approx 1$. Only the term proportional to $e^{-iqx}$ in the
second line on the rhs of Eq.(\ref{gkxtp}) contributes to the gain.
The $x$-integration in Eq.(\ref{nktres}) is trivial and one obtains
\begin{equation}
  \label{nktp21}
  \left< n_k \right>^{(2)} (t) = \left( \frac{2 \pi}{L} \right)^2 
  \sum_{i<j} {}  
  \left| \gamma_{k-k_i^0}(t) - \gamma_{k-k_j^0} (t) \right|^2 +
  O(\gamma^3).
\end{equation}

There are two types of terms missing in Eq.(\ref{gxtp}) in order to
make $G_{N-1}^{(i)}$ correct to order $g^4$. They are due to single
pair permutations in Eq.(\ref{gxt})

\begin{equation}
  \label{gxtp2}
  G_{N-1}^{(i),(2)} = - \left( G_0(x,t) \right)^{N-3} \sum_{(m \neq n)
    \neq i} \left( G(k_n^0 - k_m^0,x,t) - G(k_i^0-k_m^0,x,t) \right)
  G(k_m^0-k_n^0,x,t). 
\end{equation}

Expanding this expression to order $\gamma^2$ and adding it to the
approximation Eq.(\ref{gxtp}) for $G_{N-1}^{(i)}$ one obtains for $k$
larger than the initial momenta

\begin{equation}
  \label{nktp2}
  \left< n_k \right>^{(2)} (t) = \left( \frac{2 \pi}{L} \right)^2 
  \sum_{i<j} {} '
  \left| \gamma_{k-k_i^0}(t) - \gamma_{k-k_j^0} (t) \right|^2 +
  O(\gamma^3)
\end{equation}

\noindent where the prime indicates that only the pairs $(i,j)$
contribute to the sum for which $k_i^0+k_j^0-k$ is different from the
other initial momenta.  This result which correctly incorporates the
Pauli blocking can also quite easily be derived in second order
time-dependent perturbation theory for the initial state (see appendix
\ref{gainpeak}) and it is very helpful for the discussion of the
results presented in the next figures.

In the following we present detailed numerical results for finite
systems using the exact expression for $G_{N-1}^{(i)}(x,t)$ as well as
the approximation presented in Eq.(\ref{gxtp}) and compare the results
with the corresponding ones for distinguishable particles with the
same initial momenta.

In Fig.\ref{6exc} the results for the distribution function of three
different systems are compared at time $\omega_0 t = 50$ with the
coupling $\alpha = 2\pi 10^{-3}$. In the first system there are six
indistinguishable particles (fermions) initially occupying adjacent
momentum states from $-5 \Delta q$ to zero.  In the second system the
total momentum distribution for six distinguishable particles
initially occupying the same states is presented. The third system
corresponds to two fermions in adjacent momentum states with the same
dimensionless density $N/(Lq_B)$ as the six fermion system.  For
initial momenta where all the particles can relax into the lower
resonant momentum state by emitting a phonon with momentum $q_B$ the
difference between the fermions and the distinguishable particles is
generally small in the loss part of the distribution up to
intermediate times as can be seen in Fig.\ref{6exc}.  The occupation
numbers in the region of the gain peak ($k=q_B$), however, differ
roughly by a factor of two. For shorter times the deviations are more
drastic.  In appendix \ref{gainpeak} the occupation numbers in the
gain region are calculated in second order time-dependent perturbation
theory for the initial state for fermions and distinguishable
particles and the results derived there can account very well for this
difference.  The physical content of Eq.(\ref{nktgainacp}) (agreeing
with Eq.(\ref{nktp2})) and Eq.(\ref{nktgainac}) is that a particle can
be scattered into a higher momentum state with an amplitude
$\gamma_q(t)$ by absorbing a phonon with momentum $q$ which was
emitted previously by a different particle. This process has an
exchange process where the initial and the final state are the same
but the role of the two particles as an emitter and an absorber of the
phonon are exchanged. The phonons exchanged in the two processes are
different but this is irrelevant for the calculation of the fermionic
momentum distribution.  For fermions the occupation probability for a
momentum state in the gain region is the absolute square of the sum of
the amplitudes for the two processes which have a relative minus sign.
For distinguishable particles on the other hand the two processes can
be distinguished and the probability is the sum of the absolute square
of the amplitudes for the individual processes. In Fig.\ref{6exc} all
the initially occupied momentum states are very close and consequently
all the corresponding exchange processes have a similar amplitude
leading to the reduction of the weight in the gain peak for the
fermions. A more detailed discussion of $\gamma_q(t)$ is given in
appendix \ref{gainpeak} and it shows that the shorter the time the
broader are the features of the amplitude and therefore the
cancellation in Eq.(\ref{nktp2}).

The interference effect contained in
Eq.(\ref{nktp2}) is not properly described by the usual quantum
kinetic equations \cite{zimmerpss159,haugprb47} in which the collision
term is treated in the Born approximation but requires an improved
treatment \cite{wfs}.
 
The agreement of the two fermion system with the corresponding six
particle system is very good - also for larger times.
The momentum grid of the smaller system is different but the results
for its momentum states nearly always agree
with the results for the larger system.
The approximation presented in Eq.(\ref{gxtp}) agrees with the exact
solution within the plotting accuracy.

In Fig.\ref{6excsc} the distributions of the same six particle systems
as in Fig.\ref{6exc} are presented at time $\omega_0 t = 50$ but with
the coupling strength $\alpha = 2\pi 10^{-2}$.  Consequently the
relaxation process is more advanced and the peaks are smoother with a
larger weight in the states between them.  The approximation
(Eq.(\ref{gxtp})) for the six indistinguishable particles -
not shown in the figure for its clarity - gives 
values slightly too large for the central excitations and phonon peaks
smaller than $-7 q_B$ and larger than $2 q_B$.

In Fig.\ref{33exc} the results are shown for the same systems at the
same time as in Fig.\ref{6excsc} but again with the weaker coupling
$\alpha = 2 \pi 10^{-3}$ and different initial momenta.  Here the zero
momentum state and the next two lower states are occupied and the
second group of three particles has momenta shifted by $-q_B$.
Therefore in the six fermion system the three particles with larger
momenta cannot relax by emitting a resonant phonon. The
distinguishable particles of course are not subject to this blocking
and in contrast to Fig.\ref{6exc} one can clearly see the large
difference between the two systems in the negative momentum peaks.
However the difference in the first gain peak is not as large as in
Fig.\ref{6exc}. For an explanation Eq.(\ref{nktp2}) is useful again.
If one particle
of the smaller and larger momentum group interact by phonon exchange
the corresponding exchange process has a very different amplitude.  In
the one process a phonon with momentum $q_B$ will be absorbed and in
the exchange process a phonon with momentum $2 q_B$.  Consequently the
contributions of the two processes do not cancel each other as much as
in Fig.\ref{6exc}. On the other hand processes involving two particles
from the larger momentum group only do not contribute to the resonant
states of the gain peak at all but the exclusion of these processes
cannot compensate the other effect and the difference
between the fermions and the distinguishable particles is reduced.

\section{summary}
\label{summary}

We have presented the exact analytical result for the time dependence
of the electronic momentum distribution for a model with a linear
electronic energy dispersion for a translationally invariant initial
state with $N$ electrons and thermal phonons at temperature $T_{ph}$.
In order to show as clearly as possible the difference between the
$N=1$ ``polaron case'' and the new effects occurring for $N>1$ we have
presented numerical results for $T_{ph}=0$ only. While for
distinguishable particles the influence function can be simply
evaluated also for a finite {\it density} of particles in the
thermodynamic limit, for fermions numerical results were presented
only for finite particle number and system size. As shown in
Fig.\ref{6exc} for not too large times finite systems with different
$N$ but the same {\it density} are very similar, which indicates that
going to the thermodynamic limit is not very important. Apart from the
obvious effects of Pauli-blocking interference effects play an
important role which are not incorporated in quantum kinetic equations
which treat the scattering term in the Born approximation
\cite{zimmerpss159} and \cite{haugprb47}. As interference effects are
very important to correctly describe the gain process in the short
time limit our exact results can serve as an excellent testing ground
for improved quantum kinetic equations.

\acknowledgements{ The authors would like to thank V.\ Meden and E.\ 
  Runge for the collaboration at early stages of this work and J.\ 
  Fricke and R.\ Zimmermann for stimulating discussions. One of us
  (C.\ W.\ ) would like to thank the Deutsche Forschungsgemeinschaft
  (SFB 345 ``Festk\"orper weit weg vom Gleichgewicht'') for financial
  support.}

\appendix 
\section{}
\label{exint}

In this appendix we present analytical results for various integrals
appearing in Eqs.(\ref{nk1tres}) and (\ref{gnxt}) for the model with a
$q$-independent coupling function $g^{(2)}(q) \equiv g$ and $\omega_q
\equiv \omega_0$. We first consider the integral

\begin{equation}
  \label{ixt}
  I(x,t) \equiv \int_{-\infty}^{\infty} dx \left( \frac{\sin{\left[
      \left( \omega_0-vq\right) t/2 \right]}}
      {\left( \omega_0 -vq \right) /2} \right)^2 e^{iqx}
\end{equation}

\noindent which is already needed for the $N=1$ ``polaron
case''. Changing the integration variable it can be written as

\begin{equation}
  \label{ixts}
  I(x,t) = \frac{2t}{v} e^{iq_B x} \int_{-\infty}^{\infty} du
  \frac{ \sin^2(u)}{u^2} e^{-iuy}
\end{equation}

\noindent with $y=2x/(vt)$. Using contour integration 
one obtains for $t>0$

\begin{equation}
  \label{ixtres}
  I(x,t) = \frac{2\pi}{v^2} e^{iq_B x} \left(vt-|x| \right)
  \Theta(vt-|x|), 
\end{equation}

\noindent where $\theta (x)$ is the step function. Similarly the function

\begin{equation}
  \label{jxt}
  J(x,t) = \int_{-\infty}^{\infty} dq \frac{1}{\left(\omega_0- v q \right)^2}
  \left( 1+i\left(\omega_0-v q\right) t -e^{i(\omega_q-v q) t}
  \right) e^{iqx}
\end{equation}

\noindent which is needed in the calculation of $G_0(x,t)$ in
Eq.(\ref{gnxt}) can be evaluated by contour integration, too. For
$t>0$ one has

\begin{equation}
  \label{jxtres}
  J(x,t) = \frac{2\pi}{v^2} e^{iq_B x} (vt - x)
  \Theta(x)\Theta(vt-x) .
\end{equation}

With these results the $q-$integration of the functions in the
integrand of the momentum distribution Eq.(\ref{nktip}) can be
performed analytically for $g^{(2)}(q)$. As a function of the
dimensionless momentum variable $\tilde{k} \equiv (k-k_1^0)/q_B$ one
obtains for $\tilde{n}(\tilde{k},\tau) \equiv q_B n(k,t)$

\begin{equation}
  \label{nktauip}
  \tilde{n}(\tilde{k},\tau) = \int_{- \infty}^{\infty}
  \frac{du}{2 \pi} e^{i \tilde{k}u} e^{\tilde{b}(u,\tau) + \tilde{n}_p
  \tilde{c}(u, \tau)}
\end{equation}

\noindent where $\tilde{b}$ and $\tilde{c}$ for $\tau > 0$ are given
by

\begin{equation}
  \label{butaui}
  \tilde{b}(u,\tau) = - \alpha \tau \left( 1+2n_B(\omega_0) \right) +
  \alpha \left[ e^{iu} \left( 1+n_B(\omega_0) \right) + e^{-iu}
  n_B(\omega_0) \right] (\tau - |u|) \Theta (\tau - |u|)
\end{equation}

\noindent and

\begin{eqnarray}
  \label{cutaui}
  \tilde{c}(u,\tau) & = & \int_{-\infty}^{\infty} dv \Bigl[ \exp \Bigl\{
  -2i \alpha \bigl[ ( \tau-(v-u)) \sin(v-u) \Theta(v-u)
  \Theta(\tau -(v-u)) \bigr. \Bigr. \Bigr. \nonumber \\
  && \Bigl. \Bigl. \bigl. - ( \tau -v) \sin (v) \Theta(v) \Theta(\tau-v)
  \bigr] \Bigr\} -1 \Bigr].
\end{eqnarray}

The functions obey the relations $\tilde{b}^{\ast}(u,\tau) =
\tilde{b}(-u,\tau)$ and $\tilde{c}^{\ast}(u,\tau) =
\tilde{c}(-u,\tau)$ which guarantee that $n(\tilde{k},\tau)$ is real.
The first term on the rhs of Eq.(\ref{butaui}) is equal to
$\tilde{b}(\infty ,\tau)$ which enters the weight $p_{k_1^0}(\tau)$
and $\tilde{c}(\infty,\tau)$ is given by

\begin{equation}
  \label{cinftau}
  \tilde{c}(\infty,\tau) = 2 \int_{0}^{\tau} dv \left[ \cos \left\{ 2
  \alpha (\tau-v) \sin v \right\} -1 \right]
\end{equation}

\noindent This integral cannot be performed analytically. Expanding in
powers of $\alpha$ yields

\begin{equation}
  \label{pk10tau}
  p_{k_1^0}(\tau) = \exp \left\{ -\alpha \tau \left( 1+2n_B(\omega_0)
  \right) - \alpha^2 \tilde{n}_p \left( \frac{2}{3} \tau^3 - \tau +
  \frac{1}{2} \sin 2\tau \right) + O(g^6) \right\}
\end{equation}

\noindent Due to the form of the functions $\tilde{b}(u,\tau)$ and
$\tilde{c}(u,\tau)$ the continuous part of $n(\tilde{k},\tau)$ can be
expressed as a {\it finite} integral over $u$ from zero to $\tau$

\begin{equation}
  \label{nktauicont}
  \tilde{n}_{cont}(\tilde{k},\tau) = 2 Re \left(\int_0^{\tau} \frac{du}{2
  \pi} e^{i(\tilde{k} - \tilde{k}_1^0)u} \left[ e^{\tilde{b}(u,\tau) +
  \tilde{n}_p \tilde{c}(u,\tau)} -p_{k_1^0}(\tau) \right] \right). 
\end{equation}

\section{}
\label{recursion}

In this appendix we shortly describe our numerical method to calculate
the momentum distribution for finite systems. It is a straightforward
generalization of the technique used in \cite {mfws}.  As one step of
the procedure can actually be simplified considerably compared to \cite
{mfws}, we first describe the method for the simplest case $N=1$ , the
coupling function $g^{(1)}(q)$ and the phonons initially in their
ground state. In this case the function $F_1 (x,t)$ in Eq.(\ref{f1xt})
can be expanded in a power series in $z \equiv e^{i( \frac{2
    \pi}{L})x}$

\begin{equation}
  \label{f1exp}
  F_1(x,t) = \sum_{m=0}^{\infty} F_m(t) e^{im(\frac{2 \pi}{L})x}.
\end{equation}

Recursion relations to determine the coefficients $F_m(t)$ are
discussed below. If we take $\rho^{(1)}(x)=1/L$, i.e.\ the initial
momentum of the particle to be $k_0=0$, the integration in
Eq.(\ref{nk1tres}) can be trivially performed and yields for $k_n = n
( \frac{2 \pi}{L})$

\begin{eqnarray}
  \label{nk1exp}
  \langle n_{k_n}^{(1)} \rangle (t) & = & \frac{1}{L} \int
  \sum_{m=0}^{\infty} F_m (t) e^{i(n+m) (\frac{2 \pi}{L})x} dx
  \nonumber \\ 
  &=& F_{-n}(t)
\end{eqnarray}

\noindent for $n \leq 0$ and $ \langle n_{k_n}^{(1)} \rangle (t) =0$
for $n>0$. In order to calculate the coefficients $F_m (t)$ we write

\begin{equation}
  \label{f1repg}
  F_1 (x,t) = e^{g_t(z)-g_t(1)} \equiv B e^{g_t(z)}
\end{equation}

\noindent where the function $g_t(z)$ follows from Eq.(\ref{f1xt}) as

\begin{equation}
  \label{gtz}
  g_t(z) = \sum_{n=0}^{\infty} a_n(t)z^n
\end{equation}

\noindent with $a_n(t)=(\frac{2 \pi}{L}) | f_{n(\frac{2 \pi}{L})} (t)
|^2$. We now expand $e^{g_t(z)}$ in a power series in z

\begin{equation}
  \label{ftzt}
  \tilde{F}_t(z) \equiv e^{g_t(z)} = \sum_{n=0}^{\infty} c_m(t) z^m.
\end{equation}

If we use $\tilde{F}_t'(z) = g_t'(z) \tilde{F}_t(z)$ we obtain the
recursion relations

\begin{equation}
  \label{cmt}
  c_m(t)= \frac{1}{m} \sum_{l=1}^{m} l c_{m-l} a_l(t). 
\end{equation}

Using $F_m(t)=B c_m(t)$ yields the momentum distribution for the
simplest case. Already relaxing the assumption $T_{ph}=0$ leads to a
new aspect as $F_1(x,t)$ can now be written as a product of a power
series in $z$ and a power series in $1/z$. Multiplying these series
and performing the integration as in Eq.(\ref{nk1exp}) leads to an
expression for $\langle n_{k_n} \rangle (t)$ which involves a
summation over a product of the expansion coefficients \cite {mfws}.

In order to calculate the momentum distribution for finite systems and
$N>1$ we also express the functions $G(n(\frac{2 \pi}{L}),x,t)$ as
Laurent series in $z$. As $G_{N-1}^{(i)}(x,t)$ is a sum over products
of these functions it also can be obtained as Laurent series in $z$
such that the final integration in Eq.(\ref{nktres}) can be performed
trivially.

\section{}
\label{gainpeak}

In this appendix we discuss the influence of the Pauli principle on
the weight of the first gain satellite in the framework of
time-dependent perturbation theory. This provides additional insight
in the meaning of the result to order $g^4$ (Eq.(\ref{nktp2})) which
we obtained from the exact solution Eq.(\ref{nktres}) and
Eqs.(\ref{gkxtp}) and (\ref{gxtp}). In order to shorten the
calculation and to simplify the argument we use the coupling function
$g^{(1)}(q)$ which vanishes for negative $q$ in this appendix and
assume the phonons to be initially in their groundstate $\left| 0
\right>_{ph}$. As we assume initial statistical operators
corresponding to homogeneous systems, it is sufficient to describe the
time evolution of an initial state $\left| {\bf k}_0 \right>_a$ for
$N$ fermions and a product state for distinguishable particles. In the
following we treat fermions and note the simplifications for
distinguishable particles in the end. We calculate $\left< n_k
\right>_t$ as

\begin{equation}
  \label{nktac}
  \left< n_k \right> (t) = \left< \phi_k(t) | \phi_k(t) \right>
\end{equation}

\noindent where $\left| \phi_k(t) \right> \equiv \hat{n}_k
\tilde{U}(t) \left| {\bf k}_0 \right>_a \left| 0 \right>_{ph}$ with
$\tilde{U}(t)$ the time evolution operator in the interaction
representation defined in Eq.(\ref{at}). As we are interested in the
{\it gain region} we assume that $k$ is {\it larger} than the largest
initial momentum $k_{max}^0$. For the coupling function $g^{(1)}(q)$
the first order perturbation theory for $\tilde{U}(t)$ gives no
contribution, i.e.\ the leading term is of second order in $g^{(1)}$

\begin{eqnarray}
  \label{phikt2}
  \left| \phi_k(t) \right>^{(2)} & = & -\hat{n}_k \int_0^{t} dt'
  \int_0^{t'} dt'' H_{ep}(t') H_{ep}(t'') \left| {\bf k}_0 \right>_a
  \left| 0 \right>_{ph} \nonumber \\
  & = & - \left( \frac{2 \pi}{L} \right) \hat{n}_k \sum_{q>0} \gamma_q(t)
  \rho_q^{\dagger} \rho_q \left| {\bf k}_0 \right>_a \left| 0
  \right>_{ph}. 
\end{eqnarray}

Here we have used Eq.(\ref{rhot}) and the fact that the term involving
$\rho_q \rho_{q'}$ does not contribute for $k>k_{max}^0$. The function
$\gamma_q(t)$ defined in Eq.(\ref{gam}) results from the double time
integration. Using Eq.(\ref{rho}) or the representation of $\rho_q$ in
second quantization one obtains

\begin{eqnarray}
  \label{nkrinst}
  \hat{n}_k \rho_q^{\dagger} \rho_q \left| {\bf k}_0 \right>_a =
  \sum_{i<j} \hat{n}_k \left( \left|k_1^0, \ldots, k_i^0+q, \ldots,
  k_j^0-q, \ldots, k_N^0 \right>_a \right. \nonumber \\ 
  + \left. \left|k_1^0, \ldots,
  k_i^0-q,  \ldots, k_j^0+q, \ldots, k_N^0 \right>_a \right)
\end{eqnarray}

\noindent where the initial momenta not written out are
unshifted. Note that the states on the rhs can vanish if one (or both)
of the shifted momenta coincide with the unshifted ones. This is the
``trivial'' manifestation of the Pauli blocking. A more subtle effect
of the antisymmetry of the states occurs, if we perform the
$q$-summation in Eq.(\ref{phikt2}). This yields

\begin{eqnarray}
  \label{phiktr}
  \left| \phi_k(t) \right>^{(2)} = - \left(\frac{2 \pi}{L} \right)
  \sum_{i<j} \left[ \gamma_{k-k_i^0} (t) \left| k_1^0, \ldots, k,
  \ldots, k_i^0+k_j^0-k, \ldots, k_N^0 \right>_a \right. \nonumber \\
  \left. + \gamma_{k-k_j^0}
  (t) \left| k_1^0, \ldots, k_i^0+k_j^0-k, \ldots, k, \ldots, k_N^0
  \right>_a \right] 
\end{eqnarray}

For fermions the second term on the rhs equals the first one apart
from a minus sign, while for distinguishable particles (without the
subscript ``a'') they are orthogonal. Using Eq.(\ref{nktac}) we obtain
for the fermionic occupation numbers in the gain regime

\begin{equation}
  \label{nktgainacp}
  \left< n_k \right>^{(2)} (t) = \left( \frac{2 \pi}{L} \right)^2 
  \sum_{i<j} {} ' 
  \left| \gamma_{k-k_i^0}(t) - \gamma_{k-k_j^0} (t) \right|^2 +
  O(\gamma^3) 
\end{equation}

\noindent where the prime indicates that the pair $(i,j)$ only
contributes if $k_i^0+k_j^0-k$ is different from all the other initial
momenta $k_l^0$ for $l$ different from $i$ and $j$. For $k-k_{max}$
larger than the range $\Delta k = k_{max} - k_{min}$ the prime on the
sum can be omitted. For distinguishable particles one obtains from
Eq.(\ref{phiktr})

\begin{equation}
  \label{nktgainac}
  \left< n_k \right>_{dis}^{(2)} (t) = \left( \frac{2 \pi}{L}
  \right)^2  \sum_{i<j}
  \left( \left| \gamma_{k-k_i^0}(t) \right|^2
  + \left| \gamma_{k-k_j^0} (t) \right|^2 \right)
\end{equation}

\noindent which also follows directly from the exact solution
Eq.(\ref{nk1tres}) summed over all particles and the approximation for
$G_0(x,t)$ presented in Eq.(\ref{gnxtp}).

For a more quantitative discussion of the difference between the gain
peaks from Eqs.(\ref{nktgainacp}) and (\ref{nktgainac}) it is
necessary to discuss the momentum dependence of $\gamma_q(t)$ for
different $\tau =\omega_0 t$. The real part of $\gamma_q(t)$ which is
equal to $|f_q(t)|^2$ has a peak at $q_B$, is symmetric to this
momentum and apart from weak oscillations falls off like
$1/(q-q_B)^2$.  The imaginary part of $\gamma_q(t)$ is antisymmetric
with respect to $q_B$, with peaks near $q_B$ and falls off like
$1/(q-q_B)$.  The width of the peaks in the real- and imaginary part
of $\gamma_q(t)$ is inversely proportional to $\tau$ and this leads to
the cancellation in Eq.(\ref{nktgainacp}).

\begin{figure}
\caption{Evolution of the weight of the initially occupied momentum
  state $p_{k_1^0} (\tau)$ in time $\tau=\omega_0 t$ with the coupling
  function $g^{(2)}$ and the coupling $\alpha=2 \pi 10^{-3}$.  As in
  all the following figures the calculations were performed with
  phonons initially at zero temperature and the coupling $g^{(1)}$ if
  not indicated differently.}
\label{tdexst}
\end{figure}


\begin{figure}
\caption{Momentum distribution function for systems with different
  coupling $g^{(1)}$ and $g^{(2)}$ and different initial distributions
  at time $\omega_0 t=50$. The coupling strength is always $\alpha=2
  \pi 10^{-3}$ and the particle density $\tilde{n}_p =0.001$. For the
  first two systems with an initially sharp momentum distribution,
  i.e.\ $\sigma = 0$ the continuous part of the distribution
  $\tilde{n}_{cont}$ is shown with the weight of the initial momentum
  state being $0.7295$ and $0.7280$ for the first , and second system
  respectively. The difference between the two curves is only visible
  in the energy oscillations for positive momenta.
  For the third system with an initial Gaussian
  distribution the distribution function $\tilde{n}$ (see
  (\ref{nktauip})) is shown.}
\label{distl50}
\end{figure}

\begin{figure}
  \caption{Momentum distribution function for the thermodynamic limit
    $\tilde{n}_{cont}$ with different particle densities and for a
    finite system with $9$ particles and $q_B = n_B \Delta q$ with
    $n_B=40$ ($\Delta q $ is the smallest momentum unit). The coupling
    strength is $\alpha = 2 \pi 10^{-2}$.  The particle density of the
    first two systems is the same ($\tilde{n}_p = 0.2/(2\pi )$). The
    weight of the initially excited state in the thermodynamic limit
    systems is $0.040$ for the first and $0.135$ for the last one.
    For the finite system the zero momentum component has been left
    out, having the value $0.042$.}
\label{tdlfins}
\end{figure}

\begin{figure}
  \caption{Momentum distribution function for three finite systems with
    coupling strength $\alpha = 2\pi 10^{-3}$ at time $\omega_0 t =
    50$.  For the two systems with six fermions and six
    distinguishable particles we use $n_B = 42$ and for the two
    fermion system $n_B = 14$ in order to give the
    same dimensionless particle density. The occupancies of the
    initially occupied adjacent states at zero momentum have been
    rescaled by $1/5$ to allow for a better resolution. }
\label{6exc}
\end{figure}

\begin{figure}
  \caption{The six particle systems are the same as in
    Fig.\ref{6exc} for $\omega_0 t = 50$ but with the coupling $\alpha
    = 2\pi 10^{-2}$.}
\label{6excsc}
\end{figure}

\begin{figure}
  \caption{The systems and the time are the same as in
    Fig.\ref{6excsc} but with $\alpha = 2\pi 10^{-3}$ and different
    initial momenta. Here, three adjacent momentum states at zero
    momentum and three shifted by $-q_B$ are initially occupied.  The
    occupancies of these states have been again rescaled by $1/5$.}
  \label{33exc}
\end{figure}

\pagebreak

\begin{center}
        \epsfig{file=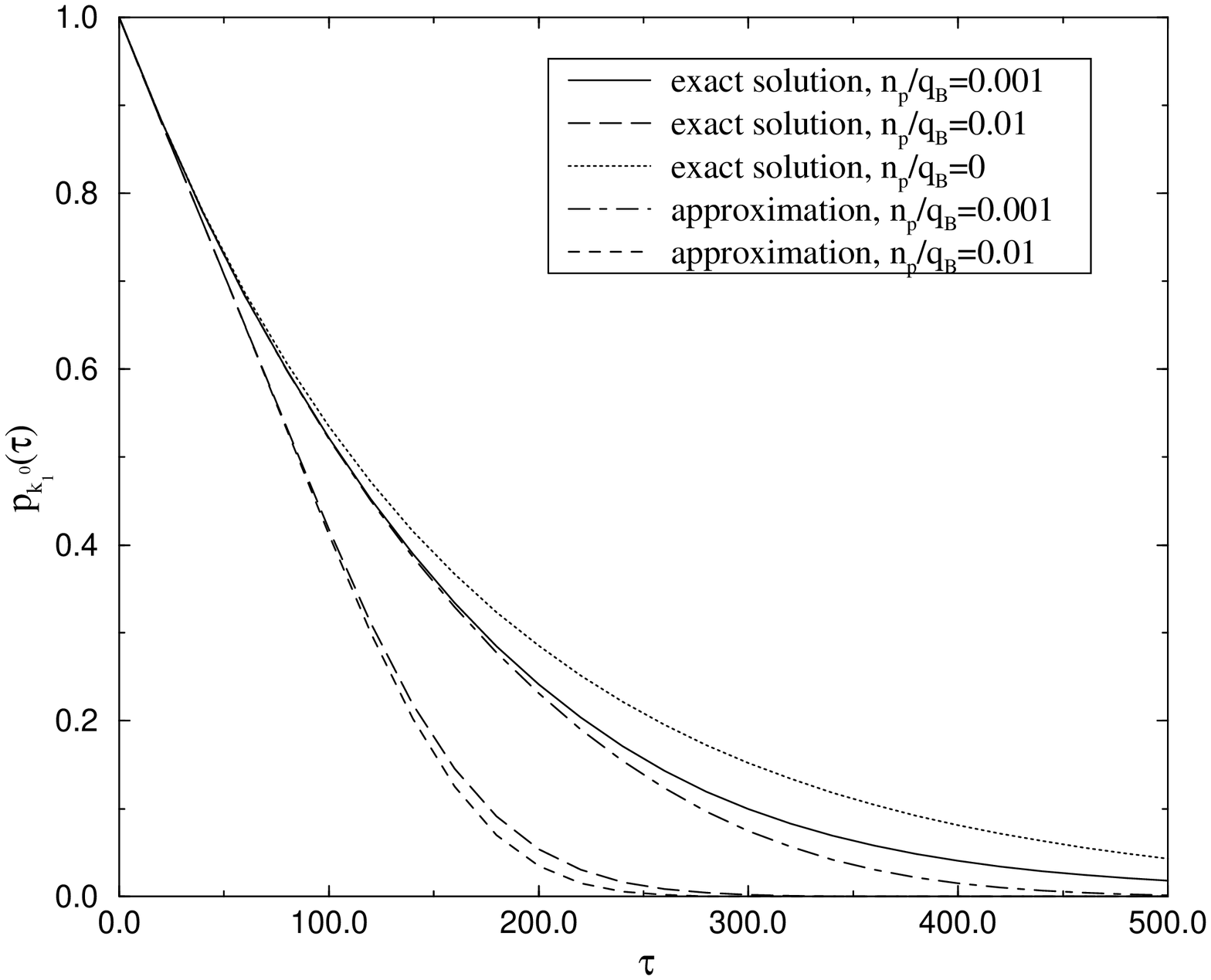,height=12cm,angle=270}
\end{center}

Figure 1

\begin{center}
        \epsfig{file=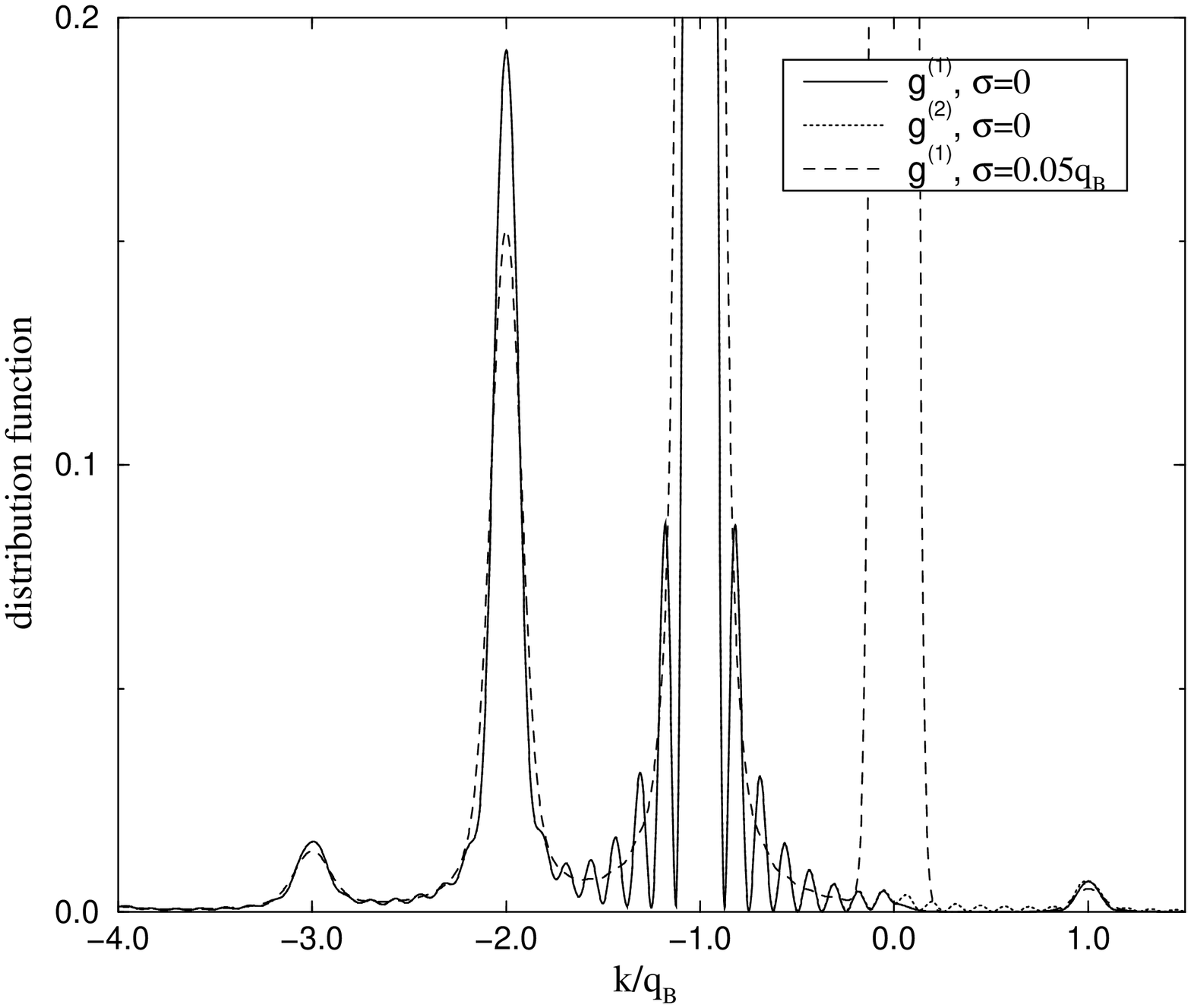,height=12cm,angle=270}
\end{center}

Figure 2

\begin{center}
        \epsfig{file=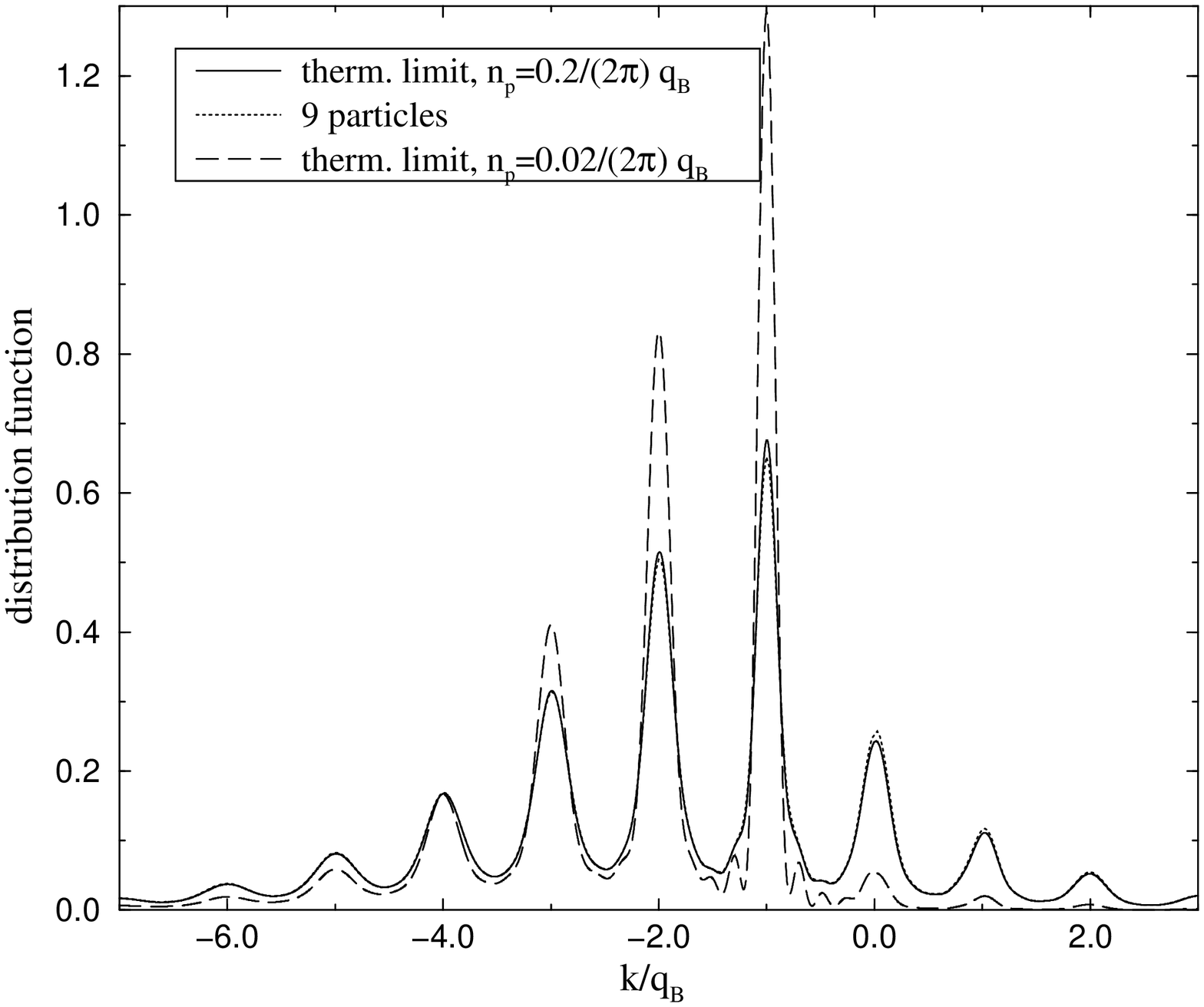,height=12cm,angle=270}
\end{center}

Figure 3

\begin{center}
        \epsfig{file=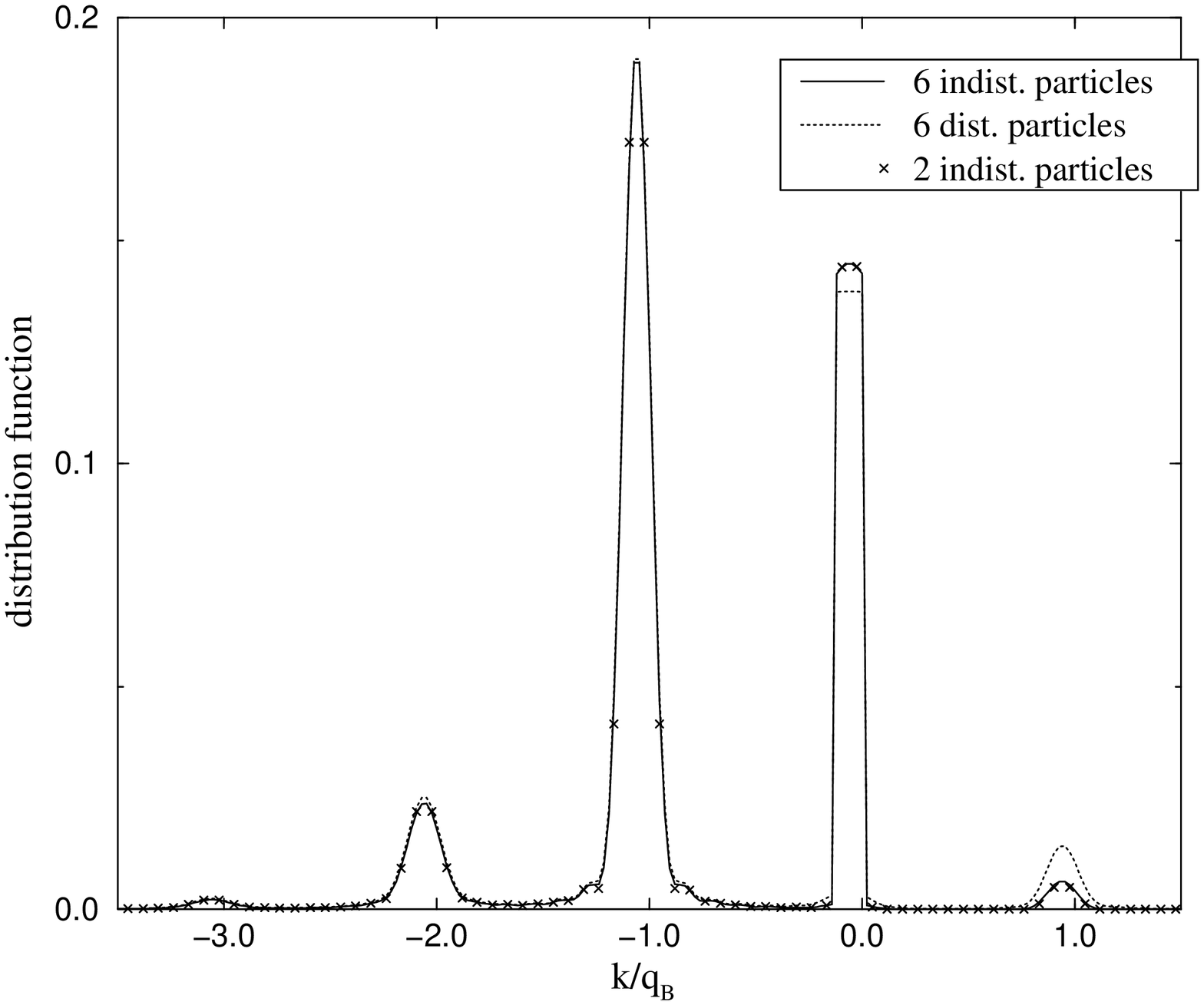,height=12cm,angle=270}
\end{center}

Figure 4

\begin{center}
        \epsfig{file=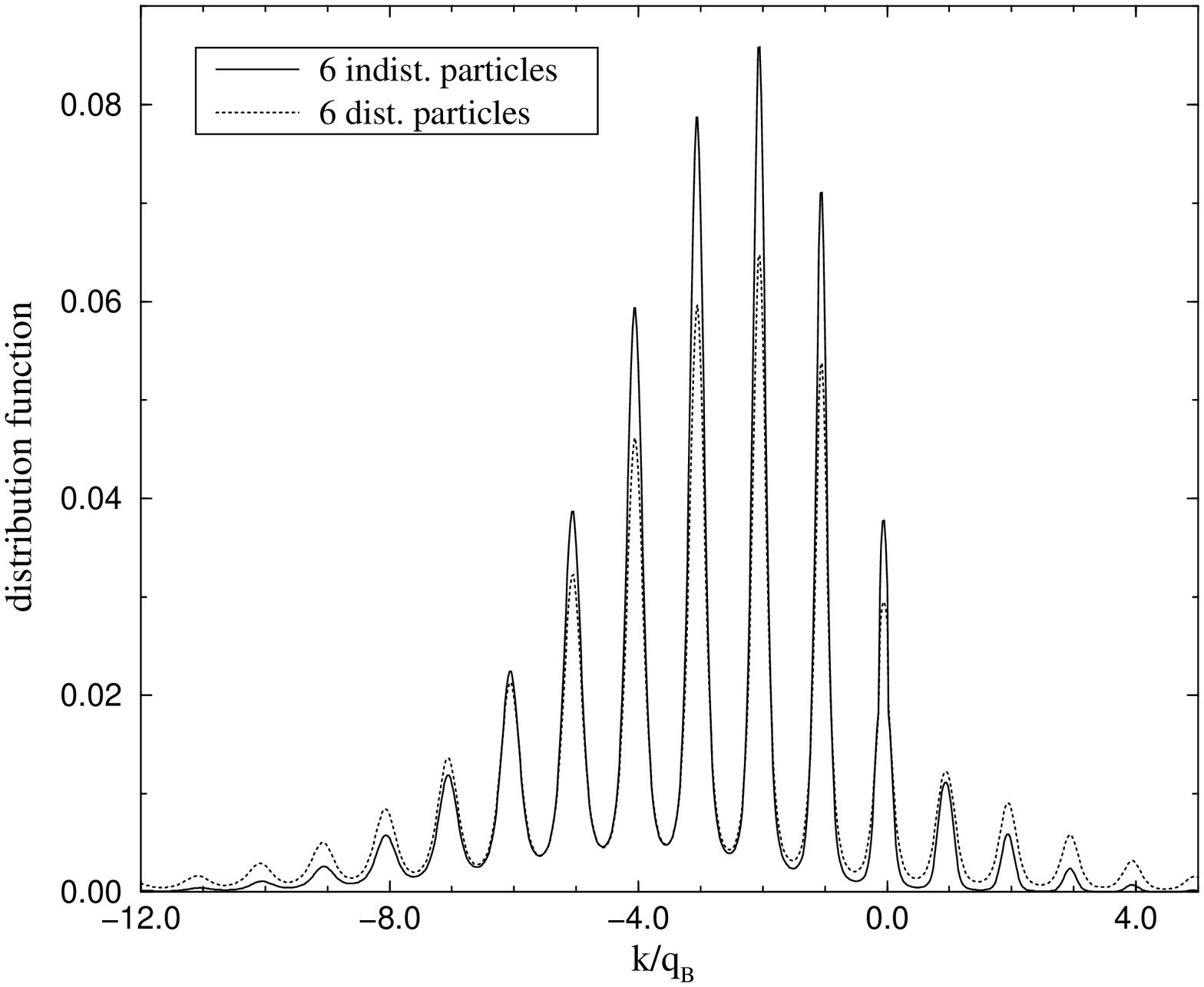,height=12cm,angle=270}
\end{center}

Figure 5

\begin{center}
        \epsfig{file=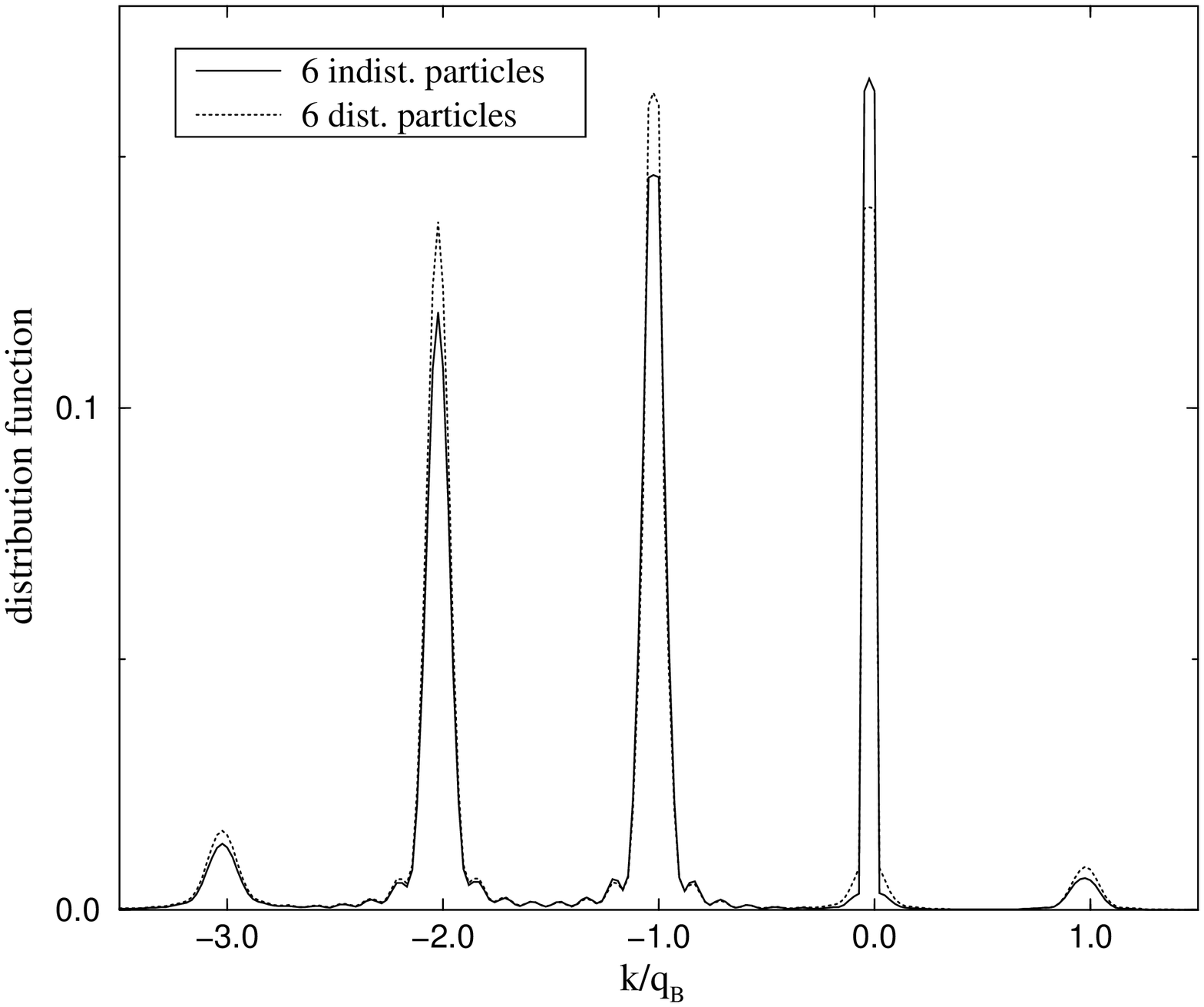,height=12cm,angle=270}
\end{center}

Figure 6

\end{document}